\documentstyle[aps,prl,epsf,amssymb]{revtex}

\def\a{\alpha}
\def\b{\beta}
\def\g{\gamma}

\def\d{\delta}
\def\D{\Delta}

\begin{document}
\draft

\twocolumn[\hsize\textwidth\columnwidth\hsize\csname 
@twocolumnfalse\endcsname

\title{Entropy-driven pumping in zeolites and ion channels}
\author{Tom Chou$^{1}$ and Detlef Lohse$^{2}$}

\address{$^{1}$Mathematics Dept., Stanford University, Stanford, CA 94305}

\address{$^{2}$Department of Applied Physics and 
J. M. Burgers Centre for Fluid Mechanics, University of Twente, 
P.O. Box 217, 7500 AE Enschede, The Netherlands}

\maketitle

\begin{abstract}

When two binary solutions are separated by a permeable barrier, the individual
species typically diffuse and mix, dissipating their chemical potential gradients. 
However, we use model  lattice simulations to show that single-file
molecular-sized channels (such biomembrane channels and zeolites) can exhibit
diffusional pumping, where one type of particle uses its entropy of mixing to
drive another up its chemical potential gradient.  Quantitative analyses of rates
and efficiencies of transport are plotted as functions of transmembrane potential,
pore length, and particle-pore interactions.  Our results qualitatively explain
recent measurements of ``negative'' osmosis and suggest new, more systematic
experiments, particularly in zeolite transport systems.


\end{abstract}

\pacs{82.65.Fr, 0.5.60.+w, 87.22.Fy, 66.30.Ny}

]

Particle transport across microscopic pores is a crucial
intermediate step in almost all biological and chemical
engineering processes.  Separation, catalysis, and drug
release all rely on controlled transport through microscopic
channels such as zeolites\cite{BARRER}.  Biological
examples include integral membrane proteins which traverse
hydrophobic, highly impermeable lipid membranes and
participate in small molecule transport\cite{ALBERTS}.  



Water transport is usually assumed to be driven by osmotic pressure
differences across pores\cite{FINK}. Solute permeability may contribute a
counterflowing solute current, which often reduces the magnitude of total
volume flux \cite{ZEUTHEN,SOLOMON}.  However, we show that for large
enough asymmetry between solvent-pore and solute-pore interactions,
volume transfer can occur in a direction opposite to that expected from
simple osmosis\cite{SU}. Here, one species of the otherwise counterflowing
pair wins and forces the second to be coflowing.  Analysis of a physically
motivated kinetic model of single-file pores shows that they can also
function as symport pumps\cite{NICHOLLS} under certain physiological
conditions, where for example B(=solutes, ions) pumps A(=water) up its
electrochemical potential gradient.  Recent osmosis experiments in various
systems \cite{ZEUTHEN,WOERMANN,MOW} have indeed revealed negative
reflection coefficients, {\it i.e.,} volume flow opposite that expected from
standard osmosis.

Since transmembrane pores have typical radii $\sim 1-3$\AA, we
use a one-dimensional exclusion model
\cite{ASEM,CHOUPRL98} to describe single-file particle flows. 
A single-file chain is divided into $N$ sections labeled $i$, each
of length $\ell\gtrsim$ molecular diameters of A and B, and
containing at most one particle of either type.  The A(B)
occupation at site $i$ is defined by $\tau_{i},\tau'_{i} \in \{0,1\}$. 
All parameters associated with A(B) will carry unprimed(primed)
notation.  The probability per unit time that an A(B) particle 
enters from the left {\it if} site $i=1$ is unoccupied is
$\a\chi_{L}(\a'\chi_{L}')$, where $\chi, \chi'$ are the reservoir
mole fractions of the particular types of particles that enter. 
Similarly, the entrance rates at $i=N$ are denoted by
$\d\chi_{R}(\d'\chi'_{R})$.  We assume $\chi'_{L} =1-\chi_{L} >
\chi'_{R}=1-\chi_{R}$ as depicted in Fig 1(a).  The A(B) exit rates
to the right(left) are denoted by $\b,\g$ and $\b',\g'$. In the chain
interior, A-type particles move to the right(left) with probability
per unit time $p(q)$ {\it only if} the adjacent site is unoccupied.
B-particles hop left(right) under the same conditions with
probability $p'(q')$.

\begin{figure}[htb] 
\begin{center} 
\leavevmode
\epsfysize=3.1in 
\epsfbox{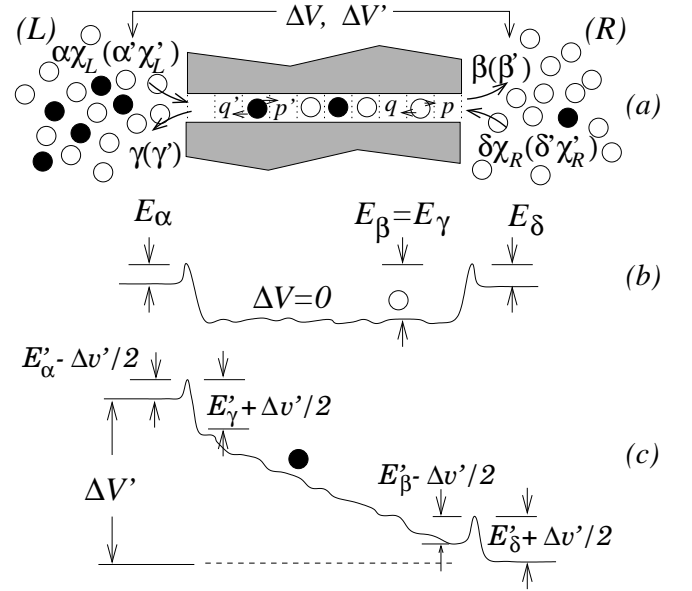} 
\end{center}
\caption{(a) A two-species 1D 
pores representing {\it e.g.} zeolites 
or biological transport channels
is divided into $N$ sections of length $\ell$. The kinetic
rate constants for A (unprimed) and B (primed)
particles are indicated. (b) and (c) depict 
energy barriers for A (open circles) and B 
(filled circles) particles. Assume A is
uncharged ($\D V \equiv V_{R}-V_{L} = 0)$, while B may be
singly charged and acted upon by a 
transmembrane potential $\D V' \equiv V'_{R}-V'_{L} <
0$.} 
\label{FIGURE1} 
\end{figure}

Rigorous results exist for the two species asymmetric exclusion
process under specific conditions\cite{ASEM}.  However,
currents for $N < 4$ can be easily found analytically for general
parameter sets by solving linear rate equations coupling all
possible particle occupancy configurations\cite{CHOUPRL99}. 
For $N > 3$, lattice simulations describing the pore dynamics
are implemented by choosing a site $i$ and finding the
instantaneous current between $i$ and $i+1$: $\hat{J}_{i} =
\hat{p}\tau_{i}(1-\tau_{i+1})(1-\tau_{i+1}')
-\hat{q}\tau_{i+1}(1-\tau_{i})(1-\tau_{i}') \in \{0,\pm 1\}$, where
$\hat{p}=1$ with probability $pdt$, and zero with probability
$(1-p)dt$ as $dt \rightarrow 0$.  Analogous expressions hold for
the distribution of $\hat{q}$.  The occupations $\tau_{i}, \tau_{i}'$
are then updated and the next particle randomly chosen.  An
analogous rule holds for $\hat{J}'(\xi')$. Boundary site kinetics
are correspondingly determined by $\a, \d, \a',\d'$, {\it e.g.}
$\hat{J}_{N}=\hat{\b}\tau_{N}-\hat{\d}(1-\tau_{N}) (1-\tau'_{N})$.
Typically $10^{9-11}$ steps are needed for $\hat{J}, \hat{J}'$ to
converge to their steady state values, $J$ and $J'$. We verified
all results by comparing simulations for both $\a'=\d'=\D V=\D
V'=0$ with exact results\cite{CHOUPRL98} and general kinetic
parameters with numerical results for $N\leq
3$\cite{CHOUPRL99}.

Since particle fluxes across microscopic channels are $<
10^9$/s \cite{FINK,ZEUTHEN}, typical fluid particles suffer $>
O(10^{3})$ ``collisions'' and relax to local thermodynamic
equilibrium (LTE) while traversing the pore.  Kinetic rates,
$\{\xi\} \equiv \{\a,\b,\g,\d,p,q\}$, under LTE take Arrhenius
forms:

\begin{equation}
\begin{array}{l}
p\simeq q \approx (v_{T}/\ell)\exp(-E_{p}/k_{B}T) \\
\b \simeq  \g \approx \left( v_{T}/\ell \right)
\exp(-E_{\beta}/k_{B}T)\\
\a,\d \simeq \a_{0}\exp(-E_{\a}/k_{B}T),
\label{PARAMETERS}
\end{array}
\end{equation}

\noindent with analogous expressions for $\{\xi'\} \equiv$
$\{\a',\b',\g',\d',p',q'\}$. We have for simplicity assumed
microscopically symmetric pores and equal hydrostatic
pressures ({\it i.e.} $\b=\g, \a=\d$ and $\b'=\g', \a'=\d'$ when
transmembrane potentials $\D V=0$ and $\D V'=0$,
respectively) and also that within a narrow pore, particles
equilibrate with the pore interior and relaxes momenta
much faster than particle positions,
implying, in the absence of external potentials, $p\simeq q,\,
p'\simeq q'$.  The hopping rates $\b, \g, p$ given in
(\ref{PARAMETERS}) represent ballistic travel times
(thermal velocity $v_{T}$ divided by section length),
weighted by internal interaction energies $E_{p}$. The
entrance energy dependent factors $(\a,\d)$ and $(\a',\d')$,
when multiplied by the relevant number fraction of entering
particles, $(\chi_{L},\chi_{R})$ and $(\chi'_{L},\chi'_{R})$,
respectively, define entrance rates into empty boundary
sites. The precise values of the prefactors $\a_{0},\a_{0}'$
will also depend on {\it equilibrium} parameters in the
reservoirs such as temperature, total number density, and
the effective area of the pore mouths.  The energy barriers
experienced during single particle transport are enumerated
in \ref{FIGURE1}(b) and (c) and may include external
potentials. If say, the B-particles have charge $ze$ and are acted
upon by a pondermotive transmembrane potential $\D V'
\neq 0$,  the energy barriers are shifted:
$E'_{\a,\b}\rightarrow E'_{\a,\b}+k_{B}T v'/2$ and
$E'_{\g,\d}\rightarrow E'_{\g,\d}-k_{B}T v'/2$, where $v' = \D
zeV'/(N+1)k_{B}T$. Thus, the kinetic rates when $v'\neq 0$ are

\begin{equation}
\begin{array}{l}
(\a',\b',p') \rightarrow (\a'_{v},\b'_{v},p'_{v})\equiv 
(\a',\b',p')e^{-v'/2} \\
(\g',\d',q') \rightarrow (\g'_{v},\d'_{v},q'_{v}) \equiv 
(\g', \d', p')e^{+v'/2} \\ 
\end{array}
\end{equation}

\noindent When A(B) is uncharged(charged), $J,\,J'$ are
computed using the parameters $\{\xi\}, \{\xi_{v}'\} =
(\a_{v}',\b_{v}',\g_{v}',\d_{v}', p_{v}', q_{v}')$, and
$\chi'_{L,R}$.  

Under isobaric, isothermal conditions, $\D E(P_{L}=P_{R})
=E_{\a}+E_{\b}-E_{\g}-E_{\d}=0$, and since
pressure fluctuations in liquid mixtures equilibrate much faster
than concentration fluctuations, the total enthalpy change per
A(B) particle translocated is $\D H(\D H') \simeq \D V(\D V')$. 
The efficiency of using species $j$ to pump $k$ can be defined
as the ratio of the average free energy gained by $k$ to the free
energy lost by $j$:

\begin{equation}
\eta_{jk} = \left[1-\theta(J^{j} \D
\mu^{j})\right]\theta(J^{k}\D\mu^{k})
{J^{k}(\chi'_{L,R})
\Delta \mu^{k} \over
J^{j}(\chi'_{L,R})\Delta\mu^{j} }.
\label{ETA}
\end{equation}

\noindent The Heaviside functions represent the
definition that flow is considered useful work only
when $j$ and $k$ are coflowing. 
Using the entropy of mixing per particle $\Delta
S' = k_{B}\ln (\chi'_{L}/\chi'_{R})$, the Gibbs free
energy change per particle, $\Delta \mu' = \Delta H' -
T\Delta S'$, is


\begin{equation}
\Delta \mu' \simeq 
k_{B}T\ln \left({\chi'_{R}\over \chi'_{L}}\right) +\Delta V',
\label{DELTAG}
\end{equation}

\noindent with an analogous expression for $\D \mu$ associated with the
transport of an A-type particle.  When concentration {\it changes}, are
not too large, higher interaction terms contributing to $\D H, \D H'$
(and hence $\D \mu, \D\mu'$) can be neglected. These correction terms 
(which are higher order polynomials in $\D \chi, \D\chi'$) can be
straightforwardly incorporated by independently measuring bulk liquid heats of mixing. 
For concreteness, we assume species $j=$ B (charged) is used to pump
$k=$ A (uncharged) and that for liquids under ambient conditions
$v_{T}/\ell \sim 1$ps$^{-1}$.  Upon setting the time scale $dt = 10$fs,
$p=q=p'=q'=0.01\ll 1$.  




\begin{figure}[htb] 
\begin{center} 
\leavevmode 
\epsfysize=1.64in
\epsfbox{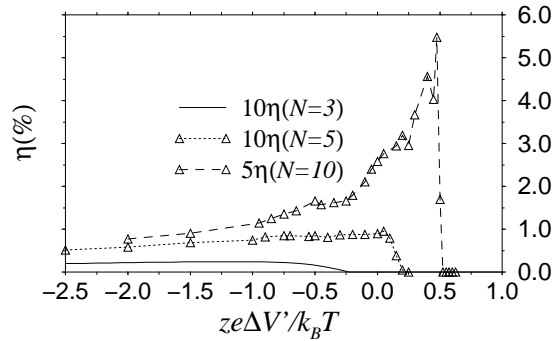} 
\end{center} 
\caption{$10\eta(N=3)$(exact, solid curves), $10\eta(N=5)$,
and $5\eta(N=10)$ as functions of $\Delta V'$.  The
corresponding parameters (for a B-attracting pore) are
$\b=\g=p=p'=q=q'=0.01$ $(dt=10\mbox{fs})$, and $\b'=\g'=
0.003$. The intrinsic entrance rates here are
$\a=\d=\a'=\d'=0.01$.  The smaller $\b'<\b, p, p'$
correspond to a pore that is more attractive to B than to
A.  Here, $\chi'_{L}/\chi'_{R} =(1-\chi_{L})/(1-\chi_{R})
=0.0036/0.0018$, correspond to a  200mM/100mM aqueous
B solution in $(L)/(R)$. The computed currents and
efficiencies are low and (numerically) noisy due to low
(physiological) concentrations $\chi'$, and entrance
probabilities $\a'\chi'$. } 
\label{FIGURE2} 
\end{figure}

Fig. \ref{FIGURE2} shows currents and efficiency as
functions of a transmembrane potential difference $\D
V'\neq 0$, for various length pores. Note that under the
physiological conditions considered, the efficiencies are
small ($\lesssim 1\%$) for pores of molecular lengths.  A
small $\vert \D \mu'\vert$ would increase efficiency via
transpore energetics; however, for too large a $\D V'$, the
B-particles are driven back against their number gradient,
and useful work precipitously vanishes.  This occurs most
easily for short channels where internal A-B interactions 
are rare; here $\D V' < 0$ is required for B to drive A uphill. 
The efficiency is nonmonotonic and has a maximum as $\D
V'$ is varied for fixed $\{\xi,\xi'\}$ and $\chi'_{L,R}$.  

For $\D V'\rightarrow -\infty$, an asymptotic form for the efficiency can
be found by assuming the B particles never hop against the
potential gradient,

\begin{equation}
\eta(\D V' \rightarrow -\infty) \sim {\a \over \a'}{k_{B}T
\over \vert \D V'\vert}
\ln\left({1-\chi'_{R}\over 1-\chi'_{L}}\right)e^{-\vert v'\vert}.
\end{equation}


Although $J'$, which is dissipating down its
electrochemical potential, generally decreases for
longer pores, for small $\D V'$,
$J(N=10)>J(N=5)>J(N=3)$ because pumping is more
efficient since there is less likelihood that a B
particle can drift through without pushing out all the A
particles ahead of it.

\begin{figure}[htb]
\begin{center}
\leavevmode
\epsfysize=2.3in
\epsfbox{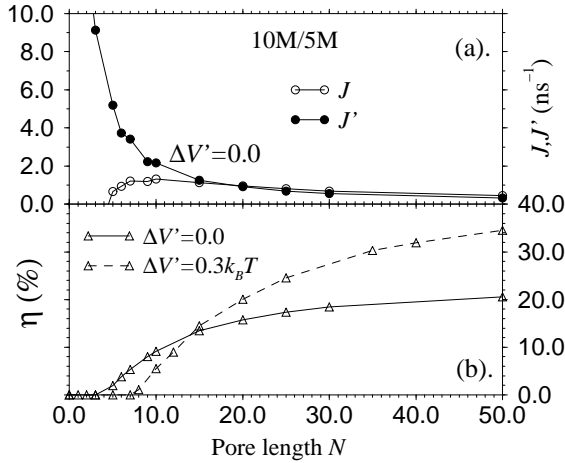}
\end{center}
\caption{Dependence of (a). currents and (b). efficiency on pore 
length $N=L/\ell$ for $ze\D V'=0.0, 0.3k_{B}T$. 
The parameters $\{\xi,\xi'\}$ used are 
the same as those in Fig. \ref{FIGURE2},
except $\chi'_{L}/\chi'_{R} = 0.18/0.09$
corresponding to a 10M/5M aqueous solution.
Although $J,J'$ generally decrease with $N$, 
efficiency increases to an asymptotic value.}
\label{FIGURE5}
\end{figure}

\noindent The particle fluctuations
across the pore that allow A and B to simply dissipate
their chemical potentials become rarer as membrane thickness or $N$
increases, enhancing efficiency as shown in Fig. 
\ref{FIGURE5}. Henceforth, unless otherwise indicated,
we treat only entropic driving, {\it i.e.,} $\D V=\D V'=0$.
Note that $\eta(N=0,1)=0$ is exact for all
parameters since $N=0$ corresponds to an infinitely
thin, noninteracting membrane, and an analytic solution
for $J(N=1) \propto \a\b(\chi_{R}'-\chi_{L}')$
\cite{CHOUPRL98}.  When $\D\mu > 0$,
$\chi_{R}'<\chi_{L}'$ and $J<0$, regardless of whether
or not solute enters or passes through the single-site pore. 
This is expected since A and B never interact within
the single-site pore for B to be able to ``ratchet'' A
through.  For larger $N$ however, we find an asymptotic
maximal efficiency defined by the kinetic parameters
$\{\xi,\xi'\}$.  As $N\rightarrow \infty$, the large
fluctuations required for net particle transport will
transfer a constant ratio of A and B particles.  This
asymptotic efficiency can be tuned by judiciously selecting $\{\xi,
\xi'\}$ which yield the desired $N\rightarrow \infty$
performance.  Note that for $\D V'=0.3$, the
$\eta(N\rightarrow \infty)$ limit is larger, but for
short pores requires larger $N$ both $J,J' > 0$ and
pumping to take effect, consistent with the results in
Fig. 2.

\begin{figure}[htb]
\begin{center}
\leavevmode
\epsfysize=2.3in
\epsfbox{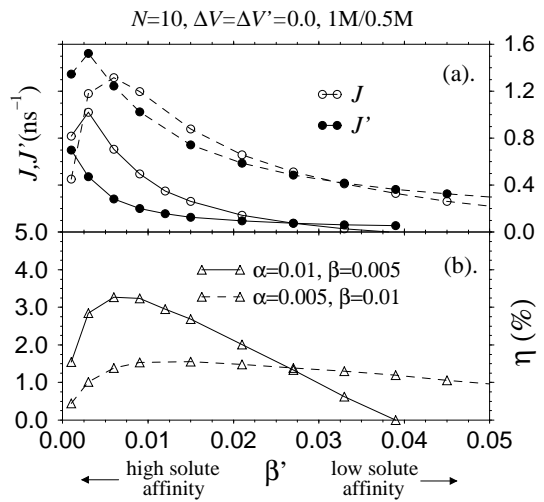}
\end{center}
\caption{(a). Currents and (b). efficiencies 
as functions of B-pore binding ($\sim 1/\b'$).
The solid curves correspond to a 
slightly A-attracting pore where $\a=0.01, 
\b=0.005$. The dashed curves 
correspond to an A-repelling pore where
$\a=0.005, \b=0.01$.}
\label{FIGURE4}
\end{figure}

Finally, performance can also be controlled by
microscopic pore-molecule interactions. For example,
recent simulations have demonstrated dynamic separation
factors between species with different pore binding
characteristics\cite{TSOTSIS}. We suggest here that
similar  molecular considerations can be used to design
pores\cite{SCIENCE} that operate as thermodynamic
pumps, without moving parts, complex cooperative
binding mechanisms\cite{NICHOLLS,HILL}, 
or external driving forces. Fig.  \ref{FIGURE4} 
shows the effect of varying B-pore
binding $\sim 1/\b'$ for two different values of A-pore
affinity $\a/\b$.

Pores that do not strongly attract (larger $\b'$\cite{EXPLAIN}) ,
or that repel solutes (smaller $\a'$) have lower B particle
average occupations $\tau'$, and allow solvent (A-particles) to
more likely pass from $(R)$ to $(L)$, decreasing efficiency. 
However, at very small $\b'$ (highly attracting pores), $\tau'$
increases to the point where it fouls the pore and decreases $J$
relative to $J'$, also decreasing efficiency.  Thus, there exists an
intermediate B-pore affinity, which yields a maximum efficiency.
Depending on A-pore affinity (dashed {\it vs.} solid curves),
currents {\it and} efficiencies can be simultaneously increased
({\it e.g.} by making pores slightly A-repelling when
$0.0028\lesssim \b' \lesssim 0.004$).  These behavior are a
consequence of the intrinsic nonlinearities and are not present
in the phenomenological linear Onsager limit \cite{HILL}.  

The above results are consistent with, and provide a microscopic
theoretical framework for recent anomalous or ``negative osmosis''
experiments\cite{WOERMANN}.  These measurements show negative
osmosis only when the chemical structure of cation exchange membranes
was modified by adding $-$CH$_{2}-$ functional groups. Thus presumably
makes the pores smaller and enhances the likelihood of finite particle size
exclusion.  More definitively, negative osmosis was measured for a window
of fixed membrane charge ($-$CH$_{2}-$SO$_{3}^{-2}$) density.  Tuning
these fixed counterion charges is equivalent to tuning $\beta'$ for
positively charged solutes (solutes which exhibited anomalous osmosis
were Ca$^{+2}$, Ba$^{+2}$, Sr$^{+2}$, and not H$^{+}$,
Na$^{+}$)\cite{WOERMANN}.  Figure \ref{FIGURE4} shows a maximal
currents and pumping efficiency as a function of $\beta'$ and matches these
experimental findings.

There is also evidence for biological manifestations of diffusional
pumping\cite{ZEUTHEN,MOW,LOO96}. Recent experiments show that water
transport is coupled to Na$^{+}$-glucose \cite{LOO96} and KCl
\cite{ZEUTHEN} transport. The negative osmotic reflection coefficient
measurements across {\it Necturus} gallbladder epithelia \cite{ZEUTHEN} in
particular have eluded explanation, although it is conjectured that separate
KCl and water transporters must be near each other in the membrane and
coupled\cite{ZEUTHEN}. However, we have demonstrated how a single
simple pore can exchange free energy (even in the absence of direct forces
such as transmembrane potentials) between two components and utilize
entropy to perform work.  This implies that under certain conditions,
common membrane channels can mimic symport pumps \cite{SU}, which
are conventionally thought of as more complicated shuttling proteins
\cite{ALBERTS,NICHOLLS,HILL}.  

Biological cell membrane channels have sizes that limit diffusional pumping
efficiencies, particularly at physiological solute concentrations (Fig. 2). 
However, the flow rates achievable by simple pores are higher than those
of shuttle enzymes and may be a viable mechanism in cellular volume
control, or whenever high fluxes are desired. The ubiquity of membrane
channels that conduct water \cite{LOO96,ZEUTHEN}, and are leaky to
certain solutes \cite{FINK,ZEUTHEN,SOLOMON}, suggests the mechanisms
presented should be considered when interpreting ``negative'' osmosis and
coupled transport experiments.  

We have not treated ``slippage,'' or incomplete coupling processes defined
by $\circ \bullet \stackrel{s} {\leftrightharpoons}\bullet \circ$ which may
occur in wider channels and decrease efficiencies.  Similarly, attractive
interactions between A-A, A-B, and B-B can also affect performance either
way. More accurate molecular dynamics or Monte-Carlo simulations may
reveal further details of diffusion pumping. Systematic measurements,
especially on more controllable artificial membrane systems where a wider
range of the parameters we have considered can be explored, may
eventually reveal secrets of more complicated bioenergetics.



TC was supported by The Wellcome Trust and The National
Science Foundation (DMS-98-04780).  DL acknowledges
support from the DFG through grant Lo556/3-1.  We have
benefited from discussions with  E. D. Siggia, T. J.  Pedley, A. E.
Hill, and S. B. Hladky.  We thank T-H Her for access to
computing facilities.



\begin{references}
\bibitem{EXPLAIN} Attractive A-pore affinities are
measured by $\b^{-1}$ since $E_{\b}$ is large and
$\b\sim \exp(-E_{\b}/k_{B}T)$ is small. As the pore is
made less attractive and eventually repulsive,
$E_{\b}\rightarrow 0$, $\b$ increases and saturates to
$\b_{0}$; eventually, $\a$ decreases as $E_{\a}$
increases (see Fig. \ref{FIGURE1}(b)). B-pore
affinities are similarly defined.
\end{references}
\end{document}